\newcommand{\be}{\begin{equation}}\newcommand{\ee}{\end{equation}}
\newcommand{\bea}{\begin{eqnarray}}\newcommand{\eea}{\end{eqnarray}}
\newcommand{\nn}{\nonumber}\newcommand{\p}[1]{(\ref{#1})}
 \newcommand{\lb}[1]{\label{#1}}
\newcommand\s{\scriptscriptstyle}
\newcommand\di{\displaystyle}

\newcommand\cD{{\cal D}}
\newcommand\cE{{\cal E}}

\newcommand\cN{{\cal N}}

\newcommand\T{\mbox{Tr}\,}

\newcommand\tdap{\theta^{\dal +}}
\newcommand\tdgp{\theta^{\dga +}}
\newcommand\tdam{\theta^{\dal -}}
\newcommand\tdgm{\theta^{\dga -}}

\newcommand{\bm}{{\bf m}}
\newcommand{\bn}{{\bf n}}

\newcommand\ab{{\alpha\beta}}

\newcommand\ada{{\alpha\dot{\alpha}}}
\newcommand\adb{{\alpha\dot{\beta}}}
\newcommand\bda{{\beta\dot{\alpha}}}

\newcommand\diag{{\di\alpha\gamma}}

\newcommand\dibg{{\di\beta\gamma}}
\newcommand\diab{{\di\alpha\beta}}

\newcommand\dibr{{\di\beta\rho}}

\newcommand\disb{{\di\sigma\beta}}

\newcommand\diar{{\di\alpha\rho}}

\newcommand\diam{{\di\alpha\mu}}
\newcommand\dinr{{\di\nu\rho}}

\newcommand\dian{{\di\alpha\nu}}
\newcommand\dirm{{\di\rho\mu}}

\newcommand\disa{{\di\sigma\alpha}}

\newcommand\digs{{\di\gamma\sigma}}
\newcommand\disg{{\di\sigma\gamma}}

\newcommand\dimn{{\di\mu\nu}}

\newcommand\dirs{{\di\rho\sigma}}

\newcommand\wxar{\widehat{x}^\diar}
\newcommand\wxag{\widehat{x}^\diag}

\newcommand\wpag{\widehat{\partial}_\diag}

\newcommand\pdag{\partial_\diag}
\newcommand\pdar{\partial_\diar}

\newcommand\da{{\dot{\alpha}}}

\newcommand\dal{{\di\alpha}}

\newcommand\dga{{\di\gamma}}
\newcommand\dmu{{\di\mu}}
\newcommand\dnu{{\di\nu}}
\newcommand\drh{{\di\rho}}
\newcommand\dsi{{\di\sigma}}

\newcommand\A{{\s A}}
\newcommand\B{{\s B}}
\newcommand\C{{\s C}}
\newcommand\D{{\s D}}
\newcommand\R{{\s R}}
\newcommand\M{{\s M}}
\newcommand\N{{\s N}}
\newcommand\W{{\s W}}
\newcommand\Z{{\s Z}}

\newcommand{\0}{{\s 0}}
\newcommand{\2}{{\s 2}}
\newcommand{\3}{{\s 3}}
\newcommand{\4}{{\s 4}}
\newcommand{\5}{{\s 5}}
\newcommand{\6}{{\s 6}}

\newcommand{\pp}{{\s ++}}
\newcommand{\m}{{\s --}}

\newcommand{\Dp}{D^{\pp}}
\newcommand{\Dm}{D^{\m}}

\newcommand{\dpp}{\partial^{\pp}}
\newcommand{\dm}{\partial^{\m}}

\newcommand{\Vp}{V^\pp}
\newcommand{\Vm}{V^\m}

\newcommand{\Dalp}{D^+_\dal}

\newcommand{\Dalm}{D^-_\dal}

\newcommand\dza{d\zeta^{\s(-4)}}

\documentstyle[12pt]{article}

\begin{document}

\begin{titlepage}
\hfill replaced version
\vspace{3cm}

\begin{center}
{\bf HARMONIC   SUPERPOTENTIALS AND SYMMETRIES IN
 GAUGE THEORIES WITH EIGHT SUPERCHARGES}\\
\vspace{0.5cm}

{\bf Boris  Zupnik}
\footnote{On leave of absence from the Institute of Applied
Physics, Tashkent State University, Uzbekistan}\\
\vspace{0.5cm}

{\it Bogoliubov Laboratory of Theoretical Physics\\
Joint Institute for Nuclear Research\\
 Dubna, Moscow Region, 141980, Russia;\\
e-mail: zupnik@thsun1.jinr.ru}
\end{center}

\begin{abstract}
Models of interactions of $D$-dimensional hypermultiplets and
supersymmetric gauge multiplets with $\cN{=}8$ supercharges $(D{\leq} 6)$
can be formulated in the framework of harmonic superspaces. The effective
Coulomb low-energy action for $D{=}5$ includes the free and Chern-Simons
terms. We consider also the non-Abelian superfield $D{=}5$ Chern-Simons
action. The biharmonic $D{=}3,\cN{=}8$ superspace is introduced for a
description of $l$ and $r$  supermultiplets and the mirror symmetry. The
$D{=}2,(4,4)$ gauge theory and hypermultiplet interactions are considered
in the triharmonic superspace. Constraints for $D{=}1,\cN{=}8$
supermultiplets are solved with the help of the $SU(2){\times}Spin(5)$
harmonics. Effective gauge actions in the full $D{\leq}3,\cN{=}8$
superspaces contain constrained (harmonic) superpotentials satisfying the
$(6{-}D)$ Laplace equations for the gauge group $U(1)$ or corresponding
$(6{-}D)p$-dimensional equations for the gauge groups $[U(1)]^p$.
Generalized harmonic representations of superpotentials connect equivalent
superfield structures of these theories in the full and analytic
superspaces. The harmonic approach simplifies the proofs of
non-renormalization theorems.
\end{abstract}


\end{titlepage}

\section{\lb{A} Introduction}

The harmonic superspace $(HS)$ has firstly been introduced for the
off-shell description of matter, gauge and supergravity superfield
theories with the manifest $D{=}4, N_\4{=}2$ supersymmetry \cite{GIK1,
GIOS2}. The $SU(2)/U(1)$ harmonics $u^\pm_i$ and corresponding harmonic
derivatives $\dpp,~\dm$ and $\partial^0$ are used for the consistent
solution of the superfield constraints in the $ N_\4{=}2$ superspace. The
basic relations for the harmonics are
\bea
&& [\dpp,\dm]=\partial^{\0}~,\quad [\partial^{\0},\partial^{\s\pm\pm}]=
\pm 2\partial^{\s\pm\pm}~, \lb{A1}\\
 && \dpp\;u^+_i=0,\quad \dpp\;u^-_i=u^+_i,\quad \partial^{\0}u^\pm_i=
\pm u^\pm_i~,\lb{A2} \\
  &&\dm\;u^-_i=0,\quad \dm\;\;u^+_i=u^-_i~. \lb{A3}
\eea

The $HS$ approach has  also been applied to consistently describe
hypermultiplets and vector multiplet in $D{=}6, N_\6{=}1$ supersymmetry
 \cite{Z1,HSW}. It is convenient to use the total number of supercharges
$\cN$ for the classification of all these models in different dimensions
$D$ instead of the number of  spinor representations for supercharges
$N_\D$. Let us review briefly the basic aspects of the $D{=}6, \cN{=}8$
harmonic gauge theory. The harmonics $u^\pm_i$ are used to
construct  the analytic $6D$ coordinates $\zeta{=}(\wxar_\A,~\tdap)$
and the additional spinor coordinate $\tdam$, where $\alpha,~\beta,~\rho
\ldots$ are the 4-spinor indices of the $(1,0)$ representation of the
$Spin(5,1)$ group and $\theta^{\pm\dal}{=}u^\pm_i\theta^{i\dal}$. The
harmonized spinor derivatives and harmonic derivatives have the following
form in these coordinates:
\bea
&&\Dalp =\partial^+_\dal ~,\quad \Dalm =-\partial^-_\dal -i\tdgm \wpag ~,
\lb{A4}\\
&& \Dp=\dpp + {i\over2}\tdap \tdgp \wpag +\tdap \partial^+_\dal ~,
\lb{A5}\\
&& \Dm=\dm + {i\over2}\tdam \tdgm \wpag +\tdam \partial^-_\dal ~,\lb{A6}
\eea
where $\wpag =\partial/\partial \wxag$.

 The Grassmann analyticity condition in $HS$ is $\Dalp\,\omega{=}0$.
Superfield constraints of $D{=}6~~SYM$ in the ordinary superspace
(central basis or $CB$) are equivalent to the integrability conditions
preserving this  analyticity. The  Yang-Mills prepotential $\Vp(\zeta,u)$
in the analytic basis $(AB)$ describes the $6D$ vector multiplet
$(A_\diar,~\lambda^\dal_i,~X^{ik})$ and possesses the gauge transformation
with the analytic matrix parameter $\lambda(\zeta,u)$
\be
\delta_\lambda \Vp =\Dp \lambda +[\Vp,\lambda]=\nabla^\pp \lambda~.
\lb{A7}
\ee

The action of the $D{=}6~SYM$ theory has the form of integral over the
full superspace \cite{Z1}
\be
 S(\Vp)={1\over g^2_\6}\sum\limits^{\infty}_{n=1} \frac{(-1)^n}{n} \int
d^{\6}\!x d^{\s8}\theta du_{\s1}\ldots du_n \frac{\T\Vp(z,u_{\s1} )
\ldots \Vp(z,u_n )}{(u^+_{\s1} u^+_{\s2})\ldots (u^+_n u^+_{\s1} )}
  \label{A8}
\ee
where $g_\6$ is the coupling constant of  dimension $d{=}1$.
 The harmonic distribution $1/(u^+_{\s1}u^+_{\s2})$ \cite{GIOS2}
satisfies the relation
\be
 \dpp_{\s1} \frac{1}{(u^+_{\s1}u^+_{\s2})}=
 \delta^{\s(1,-1)}(u_{\s1},u_{\s2})~,\lb{distr}
\ee
where $\delta^{\s(1,-1)}$ is the harmonic $\delta$-function.

The gauge variation of this action
\be
\delta_\lambda S(\Vp)={1\over g^2_\6}\int d^{\6}\!x d^{\s8}\theta du\T
\nabla^\pp \lambda \Vm=-{1\over g^2_\6}\int d^{\6}\!x d^{\s8}\theta du
\T \lambda\Dm \Vp=0 \lb{A9}
\ee
 vanishes due to the analyticity of the parameter and prepotential.
 We have used here the harmonic zero-curvature equation
\be
\Dp \Vm -\Dm\Vp +[\Vp,\Vm]=0~,\lb{A10}
\ee
where $\Vm$ is the connection for the harmonic derivative $\Dm$.
Note that the superfield density of the gauge actions in the full
superspace is not invariant for any $D$ in contrast to the chiral
density of the $D{=}4$ gauge action.

Reality conditions for the harmonic connections include the special
conjugation of harmonics  preserving the $U(1)$-charges \cite{GIK1}
\be
\widetilde{u^\pm_i}=u^{i\pm}~,\qquad (V^{\s\pm\pm})^\dagger=-V^{\s\pm\pm}
~.\lb{A10b}
\ee

The physical fields of the hypermultiplet $f^i$ and $\psi_\dal$ and the
infinite number of auxiliary fields are  components of the analytic
$6D$ superfield $q^+(\zeta,u)$. The interaction of the hypermultiplet
and gauge field can be written in the analytic superspace
\be
S(q^+,\Vp)=\int \dza du \tilde{q}^+ (\Dp +\Vp)q^+ ~,\lb{A11}
\ee
where $\dza{=}d^\6\!x_\A(D^-)^4$ is the analytic measure in $HS$.

Universality of  harmonic superspaces is connected with the possibility
of constructing $\cN{=}8$ models in $D {<} 6$ by a dimensional reduction.
The $HS$ analysis of the $D{=}4$ low-energy effective actions has been
considered for the gauge superfields \cite{BBIKO} and for the
hypermultiplets \cite{IKZ}. The manifestly supersymmetric calculations in
$HS$ are in a good agreement with the basic ideas of the Seiberg-Witten
theory \cite{SW1}, however, the $HS$ geometry allows us to rewrite the
chiral-superspace Coulomb action as the integral in the full superspace
\be
S_\4=i\int d^\4\!xd^\4\!\theta {\cal F}(W) +\mbox{c.c.}=
\int  d^\4\!xd^{\s8}\!\theta du \Vp\Vm [F(W) +\mbox{c.c.}]~,\lb{A12}
\ee
where $F(W){=}-iW^{-2}{\cal F}(W)$ is the holomorphic part of the
superpotential in this representation and $W=(\bar{D}^+)^2\Vm$. We have
used the following decompositions of the chiral and full Grassmann
measures in terms of the harmonized spinor derivatives:
\be
d^{\s8}\!\theta=(D^+)^2(\bar{D}^+)^2 (D^-)^2(\bar{D}^-)^2~,
\qquad d^{\4}\!\theta=(D^+)^2 (D^-)^2~,\lb{measure}
\ee
where $(D^\pm)^2{=}(1/2)D^{\alpha\pm}D^\pm_\alpha$ and
$(\bar{D}^\pm)^2{=}(1/2)\bar{D}_{\dot{\alpha}}^\pm
\bar{D}^{\dot{\alpha}\pm}$.

It should be stressed that  the superpotential $f(W,\bar{W}){=}[F(W) +
\mbox{c.c.}]$ in the full-superspace representation satisfies
the constraints
\be
D^+_\alpha \bar{D}^+_\da  f(W,\bar{W})=0~\rightarrow~\partial_\W
\bar{\partial}_\W f(W,\bar{W})=0
~,\lb{A12b}
\ee
which follow from the gauge invariance
\bea
&&\delta_\lambda S_\4 \sim \int  d^\4\!xd^{\s8}\!\,\theta\, du\,
\lambda\,\Dm \Vp f(W,\bar{W})\nn\\
&&\sim \int  d^\4\!x (D^-)^4\, du\,\lambda\,\partial^\ada
\Vp D^+_\alpha \bar{D}^+_\da  f(W,\bar{W})=0~.\lb{gauginv}
\eea
 Representations of the action in the full, analytic and chiral
superspaces are also important for the $HS$
interpretation of the $4D$ electric-magnetic duality \cite{IZ}.

The holomorphic action $S_\4$ can be reduced to lower dimensions,
however, this reduction does not produce the general effective action.
The $\cN{=}8$ supersymmetries have some specific features for each
dimension based on differences in the structure of Lorentz groups $L_\D$,
maximum automorphism groups $R_\D$ and the set of central charges $Z_\D$.
The main result of this work is a construction of the  Coulomb effective
actions for the dimensions $D{=}1,2,3$ and 5 in the full $\cN{=}8$
superspace
\be
S_\D=\int  d^\D\!x\,d^{\s8}\!\theta\, du\, \Vp\Vm f_\D(W)~,\lb{A13}
\ee
where $f_\D(W)$ is the superpotential and $W$ is the constrained
$(6{-}D)$-component superfield strength for the $U(1)$ gauge
prepotential $\Vp$. The gauge invariance of this action implies the
$(6{-}D)$-dimensional Laplace equation for the general superpotential
\be
\Delta^w_{\s 6-D} f_\D(W)= 0~,\lb{A14}
\ee
which generalizes the $2D$-Laplace equation \p{A12}. The
$(6{-}D)$-harmonic solutions of this equation can be used for a
description of non-perturbative solutions in the $\cN{=}8$ gauge
theories. We discuss harmonic-integral representations of the $D{\leq}3$
superpotentials which allow us to construct the equivalent
analytic-superspace representations of $S_\D$.

It should be remarked that the function $f_\D$ determines
$\sigma$-model structures and interactions of the $(6{-}D)$-dimensional
scalar field with fermion and vector fields.

Renormalization theorems in this approach are connected with the
$R_\D$-invariant solutions of Eq.\p{A14}
\be
f_\D^\R(w_\D)=g^{-2}_\D + k_\D w_\D^{D-4}~,\qquad D\neq 4~,\lb{A15}
\ee
where the invariant superfield $w_\D$ can be interpreted as a length in
the $(6{-}D)$-moduli space, and $g_\D$ and $k_\D$ are coupling constants.

The effective actions of the $[U(1)]^p$ gauge theories are considered
also by these methods. The matrix superpotentials
of these theories satisfy the $(6{-}D)p$-dimensional Laplace-type
equations which are evident for $D{=}4$ and 5, and also for the
$D{\leq} 3$ superpotentials in the harmonic-integral representations.
The  harmonic structures of moduli spaces for the $D{\leq} 3,~\cN{=}8$
theories arising in connection with the equations for superpotentials
generalize  the original $SU(2)$-harmonic structure of the $D{\geq} 4,
\cN{=}8$ theories. These structures are necessary to classify various
$D{\leq} 3$ supermultiplets. All alternative Grassmann analyticities are
compatible with basic supersymmetries and reflect the rich $HS$ geometry
of these supersymmetric theories.

Sect.\ref{B} is devoted to the 5-dimensional supersymmetric gauge
theories. The $HS$ approach is natural for the perturbative and
nonperturbative analysis of these theories. The unique effective Abelian
action of the $D{=}5$  theory in $HS$ contains the free terms and the
cubic Chern-Simons terms. This uniqueness is the symmetry basis of the
quantum stability of these theories. We also construct the
non-Abelian superfield Chern-Simons term.

In Sect.\ref{C}, we consider the biharmonic superspace $(BHS)$
using  harmonics of the automorphism group $SU_l(2){\times}SU_r(2)$ in
the $D{=}3,~\cN{=}8$ models. The 3-dimensional $l$-vector multiplet
can be described in terms of the $SU_l(2)$ harmonics, however, the
$SU_r(2)$ harmonics arise in the integral representation of the
general $D{=}3$ low-energy superpotential $f_\3$. The Grassmann
$r$-analyticity generalizes  the holomorphicity in the $HS$ description
of the $3D$ low-energy actions. The $l$-analytic gauge
prepotentials and hypermultiplets have their mirror partners in the
$r$-analytic superspace.

The $D{=}2,~(4,4)$ models in the triharmonic $SU_c(2){\times}SU_l(2)
{\times}SU_r(2)$ superspace $(THS)$ \cite{IS1,IS2,Iv1} are discussed in
Sect.\ref{D}. We underline the importance of the (4,4) gauge theory and
derive the formula for the effective action in the full superspace
using the $SU_c(2)$ harmonics. The integral representation of the $D{=}2$
superpotential $f_\2$  in the $SU_l(2){\times}SU_r(2)$ harmonic space
contains the $rl$-analytic function of the primary analytic superfield.
The full-superspace effective action
is equivalent to the action in the $rl$-analytic superspace. The $c$-,
$l$- and $r$-analytic superspaces are convenient in classifying the (4,4)
representations and duality relations.

The one-dimensional models with 8 supercharges are used in Matrix theory
describing the D0-D4 brane interactions. These models have been
intensively studied in the field-component formalism and the $\cN{=}4$
superspace. An adequate superfield description of the $D{=}1,\cN{=}8$
theories requires the use of harmonics for the automorphism group
$R_{\s1}{=}SU_c(2){\times}Spin(5)$. We define the corresponding $BHS$
gauge and hypermultiplet models in Sect.\ref{E}.

Problems  of the $\cN{=}8$ gauge theories  have earlier been discussed in
the framework of the component-field formalism or the formalism with
$\cN{=}4,~ D{=}1, 2, 3$ superfields (see e.g.\cite{Se,InS,DS,DE}).
In particular, the $(6{-}D)$ Laplace equations have been considered in
the $\cN{=}4$ superfield formalism of the $\cN{=}8$ gauge theories and in
the formalism of the corresponding $\sigma$-models. Nevertheless,
it should be stressed that the manifestly covariant $HS$ approach provides
the most adequate and universal methods to solve the problems of the
$\cN{=}8$ theories in all dimensions. A short discussion of these ideas
has also been presented in \cite{Z4}.

It should be remarked that the general mathematical formalism for
low-dimensional harmonic superspaces has been considered in ref.
\cite{HL}. This approach treats $HS$ as homogeneous spaces of
corresponding complex superconformal groups. We do not discuss
the superconformal transformations in this paper and use harmonic
variables to find  covariant separations of the spinor coordinates
in the superfield theories.  Our formulations of
different harmonic superspaces are connected with the alternative
off-shell representations of low-dimensional supermultiplets, their
interactions and duality relations.

\setcounter{equation}{0}
\section{\lb{B} Five-dimensional harmonic gauge theories}

The consistent non-anomalous five-dimensional supersymmetric gauge
theories have been discussed in refs.\cite{Se,IMS}. The Coulomb phase
of these theories contains the cubic $5D$ Chern-Simons terms for the
gauge fields and cubic interaction of the scalar fields. We shall
consider the   $D{=}5,~\cN{=}8~$ $HS$-formalism which
is very natural for the perturbative and nonperturbative analysis
of quantum problems in these theories.

Let us consider firstly the harmonic superspace with the $D{=}5,~\cN{=}8$
supersymmetry. The general five-dimensional superspace has the
coordinates $z{=}(x^\bm,~\theta^\dal_i)$, where ${\bf m}$ and  $\alpha$
are the 5-vector and 4-spinor indices of the Lorentz group $L_\5{=}
SO(4,1)$, respectively, and $i$ is the 2-spinor index of the automorphism
group $R_\5{=}SU(2)$. The spinors of $L_\5$ are equivalent to the pair of
the $SL(2,C)$ spinors: $\Psi^\dal{=}(\psi^\alpha,~
\bar{\psi}^{\dot{\alpha}})$.

The invariant symplectic matrices $\Omega_\diar$ and $\Omega^\diar$ can
be constructed in terms of the $SL(2,C)$ $\varepsilon$-symbols
$$
\Omega_\diar=\left(\begin{array}{ll}
\varepsilon_{\alpha\rho}& \;\;0\\
\;\;0 & \varepsilon_{\dot{\alpha}\dot{\rho}}\\
\end{array}\right)~,\qquad \Omega_\diar \Omega^\dirs=\delta^\dsi_\dal~.
       \eqno{(2.1)}
$$
These matrices connect spinors with low and upper indices.

The antisymmetric traceless representation of the $\Gamma$-matrices
contains  the $4D$ Weyl matrices $\sigma_m$ and $\varepsilon$-symbols
$$
(\Gamma_m)_\diab=\left(\begin{array}{ll}
\;\;\;0& (\sigma_m)_\adb\\
-(\sigma_m)_\bda& \;\;\;\;0\\
\end{array}\right)~,\quad
(\Gamma_4)_\diab=\left(\begin{array}{ll}
i\varepsilon_{\alpha\beta}& \;\;0\\
\;\;0 & -i\varepsilon_{\dot{\alpha}\dot{\beta}}\\
\end{array}\right)~.             \eqno{(2.2)}
$$
\setcounter{equation}{2}
The corresponding representation of the $5D$ Clifford algebra has the
following form:
\be
(\Gamma_\bm)_\diab(\Gamma_\bn)^\dibg +(\Gamma_\bn)_\diab
(\Gamma_\bm)^\dibg=-2\delta_\dal^\dga\eta_{\bm\bn}~,\lb{B2}
\ee
where $(\Gamma_\bn)^\dibg=\Omega^\dibr\Omega^\digs(\Gamma_\bn)_\dirs$ and
$\eta_{\bm\bn}$ is the metric of the (4,1) space.

The  5-vector projector in the spinor space is
\be
(\Pi_\5)^\diag_\dirs={1\over4}(\Gamma^{\bm})^\diag
(\Gamma_{\bm})_\dirs ={1\over2}(\delta^\dal_\drh
\delta^\dga_\dsi-\delta^\dal_\dsi\delta^\dga_\drh)+
{1\over4}\Omega_\dirs\Omega^\diag~.\lb{B5}
\ee

Consider also the relations between the antisymmetric 4-spinor symbol
$\cE$ and the matrices $\Omega$ and $\Gamma$
\be
\cE_{\diar\dimn}=\Omega_\diar\Omega_\dimn + \Omega_\diam\Omega_\dinr
+\Omega_\dian\Omega_\dirm =-{1\over2}(\Gamma^{\bm})_\diar
(\Gamma_{\bm})_\dimn+{1\over2}\Omega_\diar\Omega_\dimn~.
\lb{B3}
\ee

It is convenient to use the bispinor representation of the $5D$
coordinates and partial derivatives
\be
x^\diar={1\over2}(\Gamma_\bm )^\diar x^\bm~,\quad
\pdar={1\over2}(\Gamma^\bm )_\diar \partial_\bm~. \lb{B4}
\ee

The $C$-conjugation rules for the $Spin(4,1)$ objects are similar to the
corresponding rules for (1,0) spinors in the $6D$ space
\bea
&&\overline{\theta}^\dal_i\equiv \varepsilon_{ik}C^\dal_\dga
(\theta^\dga_k)^* =\theta^\dal_i~,\quad (C^2)^\dal_\dga=-
\delta^\dal_\dga~, \lb{B5b}\\
&& \overline{\Omega}_\diar=-\Omega_\diar~,\quad\overline{x}^\diar=
x^\diar~,\quad\overline{\partial}_\diar=-\pdar
~.\lb{B5c}
\eea

The basic relations between the spinor derivatives of the $D{=}5,
\cN {=}8$ superspace have the following form:
\be
\{D^k_\dal~,~D^l_\dga\}=i\varepsilon^{kl}(\pdag
+{1\over2} \Omega_\diag Z)~, \lb{B6}
\ee
where $Z$ is the real central charge. We shall consider the basic
superspace with $Z{=}0$ and introduce the central charges via the
interaction of gauge superfields  satisfying the constraints
\be
\{\nabla^k_\dal~,~\nabla^l_\dga\}=i\varepsilon^{kl}(\nabla_\diag
+{1\over2} \Omega_\diag W)~, \lb{B6d}
\ee
where $W$ is the real superfield.

The spinor $SU(2)/U(1)$ harmonics $u^\pm_i$  can be used
to construct the $R_5$-invariant $HS$  coordinates
$\zeta{=}(x_\A^\bm,~\tdap),~\tdam $, spinor derivatives $D^\pm_\dal$
and harmonic derivatives by analogy with Eqs.(\ref{A4}-\ref{A6})
\bea
&&\Dalp =\partial^+_\dal ~,\quad \Dalm =-\partial^-_\dal -i\tdgm \pdag~,
\lb{B6b}\\
&& \Dp=\dpp + {i\over2}\tdap \tdgp \pdag +\tdap \partial^+_\dal ~.
\lb{B6c}
\eea

We shall use the following notation for degrees of the spinor derivatives:
\bea
&& D^{(\pm 2)}={1\over4}D^{\dal\pm}D^\pm_\dal~,\quad D^{(\pm 2)}_\diag =
 (\Pi_\5)^\dirs_\diag D_\drh^\pm D^\pm_\dsi~,\lb{B7}\\
&& D_\dal^{(\pm 3)}=D_\dal^\pm D^{(\pm 2)}~,\quad D^{(\pm 4)}=2
D^{(\pm 2)}D^{(\pm 2)}\lb{B8}
\eea
and the important identities
\bea
&&D^{(+2)}D^{(+2)}_\diag=0~,\qquad D^{(+2)}_\diag D^{(+2)}_\dirs=
-2(\Pi_\5)_{\diag,\dirs}D^{(+4)}~,\lb{B9}\\
&&D^{(+4)}\Dm\Dm D^{(+4)} =-2\partial^\bm\partial_\bm D^{(+4)}~.\lb{B9d}
\eea

The analytic Abelian prepotential $\Vp(\zeta,u)$ describes the $5D$ vector
supermultiplet. In the $WZ$-gauge, this harmonic superfield contains the
real scalar field $\Phi$, the Maxwell field $A_\bm$, the isodoublet of
spinors $\lambda^\dal_i$ and the auxiliary isotriplet $X^{ik}$
\bea
&&\Vp_{\s WZ}=i\Theta^{(+2)}\Phi(x_\A)+\Theta^{(+2)\diar} A_\diar(x_\A)
\nn\\
&&+ \Theta^{(+2)}\theta^{\dal+}u^-_i\lambda^i_\dal(x_\A) +
i[\Theta^{(+2)}]^2 u^-_k u^-_j X^{kj}(x_\A)~,\lb{B9b}
\eea
where
\be
\Theta^{(+2)}={1\over4}\theta^{\dal+}\theta^+_\dal~,\qquad
\Theta^{(+2)\diar}=(\Pi_\5)^\diar_\dimn\theta^{\dmu+}\theta^{\dnu+}~.
\lb{B9c}
\ee

The   superfield strength of this theory can be written in terms of
the harmonic connection $\Vm(\Vp)$ (see Eqs. \p{A10} and \p{B19})
\be
W_\A= -2i D^{(+2)}\Vm~,\qquad W^\dagger_\A = W_\A~.\lb{B10}
\ee
 This superfield satisfies the following constraints:
\bea
&& \nabla^\pp W_\A=\Dp W_\A + [\Vp,W_\A]=0~,\lb{B11}\\
&&D^{(+2)}_\diar W_\A=0~,\lb{B12}
\eea
where the relations \p{A10} and \p{B9} are used.
The Abelian superfield $W_\A{=}W$ does not depend on harmonics.

The $5D~SYM$ action has the universal form \p{A8} in the full harmonic
superspace. The $SYM$ equations $D^{(+4)}\Vm{=}0$
have the vacuum Abelian solution
$v^{\s\pm\pm}{=}i\Theta^{(\pm2)} Z$ where $Z$ is the linear combination
of the
Cartan generators of the gauge group ( see the analogous $D{=}4$ solution
in ref.\cite{IKZ}). This vacuum solution spontaneously breaks the gauge
symmetry, but it conserves the $D{=}5$ supersymmetry with the central
charge and produces $BPS$ masses of the $Z$-charged fields. The harmonic
supergraphs of this theory can be constructed by the analogy with
refs.\cite{GIOS2,Z1}. We do not analyze here the one-loop contribution
to the low-energy effective action of this non-renormalizable theory,
however, it is not difficult to consider the general symmetric framework
for the description of such actions in $HS$.

Chiral superspaces are not Lorentz-covariant in the case $D{=}5$, so
one can use the full or analytic superspaces only. It is readily to
construct the most general low-energy effective $U(1)$-gauge action in
the full $ \cN{=}8$ harmonic superspace
\be
S_\5=\int d^\5\!x d^{\s8}\theta du\; \Vp \Vm [ g^{-2}_\5+k_\5 W]~,
\lb{B13}
\ee
where $g_\5$ is the coupling constant of dimension $1/2$, and $k$ is the
dimensionless constant of the $5D$ Chern-Simons interaction.
Note that the next-to-leading order effective Abelian $5D$ action can be
written via the manifestly gauge invariant function H(W), but we
do not consider these terms.

The linear superpotential $f_\5{=}g^{-2}_\5+k_\5 W$ is a solution of the 
constraints
\be
D^{\s\pm\pm} f_\5=0~,\qquad D^{(+2)}_\diar f_\5=0~,\lb{B14}
\ee
which arise from the gauge invariance of $S_\5$.

It is evident in the $HS$ approach that the unique effective action
$f_\5$ cannot be renormalized by
any consistent calculations preserving the supersymmetry
and $U(1)$-gauge symmetry.

Note that the $R_\5$ invariance of the effective action can be broken by
the Fayet-Iliopoulos term in the analytic superspace
\be
S_{\s FI}=\int \dza du\;\xi^{ik}u^+_i u^+_k \Vp~,\lb{B15}
\ee
which implies also the spontaneous breaking of supersymmetry.

The gauge-invariant Chern-Simons term for the group $[U(1)]^p$ contains
the following cubic interactions of the Abelian prepotentials
$\Vp_\B $ and corresponding constrained superfields $\Vm_\B$ and
$W_\B$
\be
\int d^\5 x d^{\s 8}\theta du\; k_{\s BCD}\Vp_\B \Vm_\C W_\D~,\lb{B16}
\ee
where $k_{\s BCD}$ are coupling constants and ${\scriptstyle B,~C,~D}
{=}1\ldots p$.

It is not difficult to construct the non-Abelian $5D$ Chern-Simons term
$S^\5_{\s CS}$ starting from the following formula of its variation
\footnote{See Note added in the replaced version.}:
\bea
&&\delta S^\5_{\s CS}=k_\5 \int d^\5\!x d^{\s8}\theta du\;\T \delta\Vp
\{\Vm,~D^{(+2)}\Vm\}\nn\\
&&=k_\5 \int \dza du\;\T \delta\Vp D^{(+4)}\{\Vm,D^{(+2)}\Vm\}
~,\lb{B17}
\eea
which guarantees the gauge invariance taking into account Eqs.
(\ref{A7},\ref{A10},\ref{B11}) and \p{B12}
\bea
&&\delta_\lambda S^\5_{\s CS}=k_\5 \int \dza du\;\T \lambda D^{(+4)}
\{D^{(+2)}\Vm,\nabla^\pp\Vm\}
\nn\\
&&=k_\5 \int \dza du\;\T \lambda D^{(+4)}\{D^{(+2)}\Vm,\Dm\Vp\}=0~.
\lb{B17b}
\eea
Stress that the analogous term with $[\Vm,D^{(+2)}\Vm]$ in $\delta 
S^\5_{\s CS}$ vanishes as an integral of the total spinor derivative.

The  scale-invariant non-Abelian action $S^\5_{\s CS}$ has the following
form: 
\be
S^\5_{\s CS}={k_\5\over3} \int d^\5\!x d^{\s8}\theta du\;\T
\Vp\{\Vm_1(\Vp),~ D^{(+2)}\Vm_1(\Vp)\}+\ldots\lb{B18}
\ee
where higher-order terms are omitted and the linear approximation of the 
perturbative solution for $\Vm$ \cite{Z2} is used
\be
\Vm_1(\Vp)=
 \int
 du_{\s1} \frac{\Vp(z,u_{\s1})}{(u^+ u^+_{\s1})^2}~.  \label{B19}
\ee

\setcounter{equation}0
\section{\lb{C} Three-dimensional  biharmonic superspace}

Three-dimensional supersymmetric gauge theories have been intensively
studied in the framework of new nonperturbative methods \cite{InS,SW,
AHIS}. Superfield description of the  $D{=}3$
theories and various applications have earlier been discussed in refs.
\cite{Sch}-\cite{BG}. Three-dimensional harmonic superspaces were
considered in refs.\cite{ZK,Z3}. The most interesting features of the
$D{=}3$ theories are connected with the Chern-Simons terms for gauge
fields and also with the mirror symmetry between vector multiplets and
hypermultiplets.

The $D{=}3,~\cN {=}8$ gauge theory can be constructed in the superspace
with the automorphism group $R_\3{=}SU_l(2){\times} SU_r(2)$. Coordinates
of the corresponding general superspace are $z{=}(x^\ab,~
\theta^\alpha_{ia})$. We use here the two-component indices $\alpha,~
\beta\ldots$ for the space-time group $SL(2,R)$, $i,~k\ldots$ for the
group $SU_l(2)$ and $a,~b\ldots$ for $SU_r(2)$, respectively.

The  relations between basic spinor derivatives are
\be
\{ D^{ka}_\alpha, D^{lb}_\beta\}= i\varepsilon^{kl}\varepsilon^{ab}
\partial_\ab + i\varepsilon^{kl}\varepsilon_\ab Z^{ab}~,\lb{C1}
\ee
where  $\partial_\ab{=}\partial/\partial x^\ab$ and $Z^{ab}$ are the
central charges which  commute with all generators exept for the
generators of $SU_r(2)$. These central charges can be interpreted as
covariantly constant Abelian gauge superfields by analogy with 
\cite{IKZ}.

The superfield constraints of the $\cN{=}8$  $SYM$ theory
in the central basis  can be written as follows:
\be
\{\nabla^{ka}_\alpha, \nabla^{lb}_\beta\}= i\varepsilon^{kl}
\varepsilon^{ab}\nabla_\ab + i\varepsilon_{\ab}\varepsilon^{kl}
W^{ab}~, \lb{C2}
\ee
where $\nabla_{\M}$ are covariant derivatives with superfield connections
and $W^{ab}$ is the  constrained superfield of the $SYM$ theory
($l$-vector supermultiplet)
\be
\nabla^{ka}_\alpha W^{bc}+\nabla^{kb}_\alpha W^{ca}+\nabla^{kc}_\alpha
W^{ab}=0~.\lb{C3}
\ee
Note that gauge transformations in $CB$ have the standard form
\be
\delta \nabla^{ka}_\alpha =[\tau(z),\nabla^{ka}_\alpha]~,\qquad
\delta W^{ab}=[\tau(z),W^{ab}]~,\lb{C3b}
\ee
where $\tau(z)$ is the matrix gauge parameter.

The simplest constraints of the $l$-hypermultiplet are
\be
\nabla^{ia}_\alpha q^k + \nabla^{ka}_\alpha q^i=0.\lb{lhyper}
\ee

It is evident that one can consider the mirror $r$-versions of superfield
constraints for the vector multiplet and hypermultiplets changing the 
roles of $SU_l(2)$ and $SU_r(2)$ indices
\bea
&& \nabla^{ia}_\alpha W^{kl}_r+\nabla^{ka}_\alpha W^{li}_r+
\nabla^{la}_\alpha W^{ik}_r=0~,\lb{rvect}\\
&& \nabla^{ia}_\alpha q^b_r + \nabla^{ib}_\alpha q^a_r=0~.\lb{rhyper}
\eea

 We shall define the general $3D$ biharmonic  superspace which has simple
properties with respect to
the exchange $l\leftrightarrow r$. The mirror symmetry connects 
$l$-vector multiplets with $r$-hypermultiplets and vice versa.

Let us consider the $l$-harmonics $u^\pm_i\equiv u^{\s(\pm1,0)}_i$ of the
group $SU_l(2)$ and the analogous $r$-harmonics $v^{\s(0,\pm1)}_a$ of the
group $SU_r(2)$. The notation of charges in $BHS$ is $(q_{\s1},q_\2)$.
The constraints of the $l$-vector multiplet and $l$-hypermultiplet can be
solved with the help of the $l$-harmonics only, so
we shall use also the  notation with the one charge for the $l$-harmonic
superspace $HS_l$ which is analogous to the $4D~HS$. The $r$-harmonic
structures arise in  the geometric description of the low-energy
self-interaction of the $l$-vector multiplets and dualities between
$l$- and $r$-type supermultiplets.

 The spinor and $l$-harmonic
derivatives have the following form in the $l$-analytic coordinates
$\zeta_l{=}(x^\ab_l, \theta^{\alpha+}_a)$ and $\theta^{\alpha-}_a$:
\bea
&&D^{b+}_\alpha=u^+_i D^{ib}_\alpha=\partial^{b+}_\alpha~,\quad
D^{b-}_\alpha=u^-_i D^{ib}_\alpha=-\partial^{b-}_\alpha+i
\theta^{\beta b-}\partial_\ab^l~,    \lb{C4}\\
&&\Dp_l=\dpp_l - {i\over2}\theta^{\alpha+}_a\theta^{\beta a+}
\partial_\ab^l + \theta^{\alpha+}_a \partial^{a+}_\alpha ~.\lb{C4e}
\eea

The following relations will be used in $HS_l$ :
\bea
&& \{D^{a+}_\alpha,~D^{b-}_\beta\}=-i\varepsilon^{ab}\partial_\ab^l,
~\quad [\Dm,~D^{a+}_\alpha]=D^{a-}_\alpha~,\lb{C4b}\\
&& D^{\s(+2)}_\ab D^{ab\s(+2)}=0~,\qquad
D^{\s(+2)}_{ab}D^{\s(+2)}_{cd}=(\varepsilon_{ac}\varepsilon_{bd}+
\varepsilon_{bc}\varepsilon_{ad})(D^+)^4~,\lb{C4d}
\eea
where
\be
 D^{\s(+2)}_\ab={1\over 2}D^{+a}_\alpha D^+_{a\beta}~,\qquad
D^{ab\s(+2)}={1\over 2}D^{a\alpha+} D^{b+}_\alpha~.\lb{C4c}
\ee

The $l$-harmonic superspace is adequate to the solution of the constraints
\p{C2}
\bea
&&u^+_i u^+_k\{\nabla^{ia}_\alpha, \nabla^{kb}_\beta\}\equiv
\{\nabla^{a+}_\alpha, \nabla^{b+}_\beta\}=0~,\lb{C5b}\\
&&\nabla^{a+}_\alpha=g^{-1}(z,u)
D^{a+}_\alpha g(z,u)~,\lb{C5}
\eea
where $g(z,u)$ is the bridge matrix \cite{GIK1}. The $l$-analytic
prepotential of the $SYM$ theory is
\bea
&&\Vp_l(\zeta_l,u)=(\Dp g)g^{-1}~,\qquad D^{a+}_\alpha \Vp_l=0~,
\lb{C6}\\
&& \delta g =\lambda g -g \tau(z)~,\qquad
\delta_\lambda \Vp_l =\Dp \lambda +[\Vp_l,\lambda]~.\lb{C6b}
\eea

The components of this superfield can be determined in the $WZ$ gauge
\bea
&&(\Vp_l)_{\s WZ}=\theta^{\alpha a+}\theta^{b+}_\alpha\Phi_{ab}(x_l) +
\theta^{\alpha a+}\theta^{\beta+}_a A_\ab(x_l)\nn\\
&&+\theta^{\alpha a+}\theta^{b+}_\alpha\theta^{\beta+}_b u^-_k
\lambda_{a\beta}^k(x_l) + i(\theta^+)^4 u^-_k u^-_j X^{kj}(x_l)~.\lb{3WZ}
\eea

The  superfield strength of the $D{=}3, \cN{=}8$ gauge theory in the
analytic basis contains the corresponding harmonic connection
$\Vm_l(\Vp_l)$
\bea
&&W^{ab}_\A=-iD^{ab\s(+2)} \Vm_l=gW^{ab}g^{-1}~,\lb{C7}\\
&& \delta_\lambda W^{ab}_\A=[\lambda,~W^{ab}_\A]~,\qquad
\delta_\lambda \Vm_l =\nabla^\m \lambda~.  \lb{C7b}
\eea
 It satisfies the following constraints:
\bea
&& D^{a+}_\alpha W^{bc}_\A+ D^{b+}_\alpha W^{ca}_\A+ D^{c+}_\alpha
W^{ab}_\A=0~,
~\Rightarrow~D^{\s(+2)}_\ab W^{bc}_\A =0~,
\lb{C8}\\
&&    D^{\s\pm\pm} W^{ab}_\A+[V^{\s\pm\pm}_l, W^{ab}_\A]=0~.\lb{C9}
\eea
which are equivalent to the $CB$-constraints \p{C3}.

The Abelian superfield $W^{ab}_\A{\equiv}W^{ab}$ does not depend on 
harmonics ($D^{\s\pm\pm} W^{ab}=0)$. The vacuum Abelian solution of the 
$SYM$ theory
\be
V^{\s\pm\pm}_\0\equiv v^{\s\pm\pm}=i\theta^{a\alpha\pm}
\theta^{b\pm}_\alpha Z_{ab}~,\qquad
(D^+)^4V^\m_\0\sim D^{\s(+2)}_{ab}W_\0^{ab}=0 \lb{C10}
\ee
is covariant with respect to the supersymmetry with central charges
$Z_{ab}$ by analogy with the case $D{=}4$ \cite{IKZ}.

The $l$-analytic hypermultiplet $q^+(\zeta_l,u)\equiv q^{\s(1,0)}$ has
the standard minimal interaction with $\Vp_l\equiv V^{\s(2,0)}$ (see 
\p{A11}). By analogy with refs.\cite{Z1,IKZ}, one can construct the free 
$HS$ propagator for this superfield in the covariantly constant 
background \p{C10}
\be
i \langle q^{+}(1)| \bar{q}^{+}(2) \rangle =
-\frac{1}{\Box^\Z_{\s1}}(D^+_{\s1})^4(D^+_{\2})^4
e^{(v_{\2} - v_{\s1})}\delta^{\s11}(z_{\s1}
- z_{\2})\frac{1}{(u^+_{\s1}u^+_{\2})^3}~,
\label{C10b}
\ee
where $\Box^\Z=\partial^\ab\partial_\ab+ Z^{ab}Z_{ab}$ and $\Dp v=v^\pp$.
 The manifestly supersymmetric perturbation theory is the important
advantage of the $HS$ approach.

One can consider also the alternative   $3D$
hypermultiplet $\omega_l(\zeta_l,u)$ and the $l$-linear multiplet
$L^{\s(2,0)}(\zeta_l,u) $
satisfying the harmonic condition $D_l^{\s(2,0)}L^{\s(2,0)}{=}0$.

We do not discuss here the harmonic-supergraph calculations of the
perturbative effective action and consider the general symmetry
framework for these constructions in the full $l$-harmonic superspace.
The low-energy $U(1)$ effective action can be expressed in terms of the
superpotential $f(W^{ab})$  which does not depend on $u^\pm$
\be
 S_\3=\int d^{\3}x d^{\s8}\theta du\; \Vp_l \Vm_l f_\3(W^{ab})~.\lb{C11}
\ee
The corresponding bosonic-field Lagrangian contains the nonlinear
$\sigma$-model interaction of the 3-component scalar field $\Phi_{ab}$
\p{3WZ} and the non-minimal interaction of this field with the Abelian
gauge field. All interactions of the field $\Phi_{ab}$ are determined
via derivatives of the function $f_\3(\Phi_{ab})$.

The gauge invariance produces the following constraint :
\bea
&&\delta_\lambda S_\3=-2\int d^{\3}x d^{\s8}\theta du \lambda \Dm
\Vp_l f_\3(W^{ab})\nn\\
&&~\sim~\int d^{\3}x (D^-)^4 du \lambda \partial^\ab \Vp_l
D^{\s(+2)}_\ab f_\3(W^{ab})=0~, \lb{C12}
\eea
where the analyticity of $\Vp_l$ and relations \p{A10} and
$d^{\s8}\theta{=}(D^-)^4(D^+)^4$ are used.

This constraint on the superpotential is equivalent to the $3D$ Laplace
equation
\be
D^{\s(+2)}_\ab f_\3(W^{ab})=0~\rightarrow~
\frac{\partial}{\partial W^{ab}}\frac{\partial}{\partial W_{ab}}
 f_\3(W^{ab})=0~.\lb{C13}
\ee

The general solution of this equation breaks the $SU_r(2)$ invariance.
The $R_\3$-invariant superpotential has the following form:
\be
f^\R_\3(w_\3)= g^{-2}_\3+ k_\3 w_\3^{-1}~,\qquad w_\3=\sqrt{W^{ab}
W_{ab}}~, \lb{C18}
\ee
where $g_\3$ is the coupling constant of dimension $d{=}-1/2$, and $k_\3$
is the dimensionless constant of the $\cN{=}8$ $WZNW$-type interaction of
the vector multiplet. This superpotential is singular at the point
$Z_{ab}{=}0$ of the moduli space. The field model is well defined in the
shifted variables $\hat{W}_{ab}{=}W_{ab}-Z_{ab}$ for nonvanishing central
charges. It should be remarked that the  superfield interactions of the
$3D$-vector multiplets with dimensionless constants (Chern-Simons terms)
have earlier been constructed for the case $\cN{=}4$ \cite{ZP1} and
$\cN{=}6$ \cite{ZK,Z3}.

The effective action \p{C18} is an example of the $D{=}3$ non-perturbative
calculation based on the $\cN{=}8$ supersymmetry and the 
$R_\3$-invariance. Stress that the $HS$ approach simplifies the proof of 
this non-renormalization theorem.

The general solution of Eq. \p{C13} can be written in the
$v$-integral harmonic representation
\be
f_\3(W^{ab})=\int dv F_\3[W^{\s(0,2)},~v^{\s(0,\pm1)}_a] ~,\qquad
W^{\s(0,2)}=v^{\s(0,1)}_a v^{\s(0,1)}_b W^{ab}
~,\lb{C14}
\ee
where $F_\3$ is an arbitrary function with $q{=}(0,0)$, and $W^{\s(0,2)}$
is the $r$-harmonic projection of the basic superfield. The proof is 
based on the following $r$-harmonic representation of the Laplace 
operator:
\bea
&&\varepsilon_{ac}
\varepsilon_{bd}\frac{\partial}{\partial W_{ab}}\frac{\partial}
{\partial W_{cd}}~\sim~\frac{\partial}{\partial W^{\s(0,2)}}
\frac{\partial}{\partial W^{\s(0,-2)}}-\Bigl(\frac{\partial}{\partial
W^{\s(0,0)}}\Bigr)^2~,\lb{C15}\\
&& \varepsilon_{ac}= v^{\s(0,1)}_a v^{\s(0,-1)}_c -v^{\s(0,1)}_c
v^{\s(0,-1)}_a~.\lb{C16}
\eea

This solution is a covariant form of the well-known integral
representation of the $3D$-harmonic functions \cite{WW}. The functions
$F_\3$ do not depend on the projections $W^{\s(0,-2)}$ and $W^{\s(0,0)}$
of the superfield $W^{ab}$ while the holomorphic functions $F(W_{\s11})$
of the chiral superfield $W_{\s11}$ are independent of the components
$W_{\s22}$ and $W_{\s12}$. The $v$-integral representation
of $F(W_{\s11})$ can depend on the 1-st harmonic component
$v^{\s(0,\pm1)}_{\s1}$ only, so the representation \p{C14} is more
general than the holomorphic representation of superpotential. We shall
show that the function $F_\3$ as well as the superfield $W^{\s(0,2)}$
satisfy  the condition of the Grassmann $r$-analyticity.

Let us introduce the following definitions and relations for the
spinor derivatives in the biharmonic superspace:
\bea
&&D^{\s(\pm1,\pm1)}_\alpha=
u^{\s(\pm1,0)}_i v^{\s(0,\pm1)}_a D^{ia}_\alpha~,\qquad
D^{\s(\pm2,\pm2)}=D^{\s(\pm1,\pm1)\alpha}D^{\s(\pm1,\pm1)}_\alpha~,
\lb{Cdhs}\\
&& [D_l^{\s(\pm2,0)},D^{\s(\mp1,\pm1)}_\alpha]=D^{\s(\pm1,\pm1)}_\alpha~,
\qquad [D_r^{\s(0,\pm2)},D^{\s(\pm1,\mp1)}_\alpha]=
D^{\s(\pm1,\pm1)}_\alpha~, \lb{Cd1}\\
&& [D_l^{\s(\pm2,0)},D^{\s(\pm1,\pm1)}_\alpha]=
 [D_r^{\s(0,\pm2)},D^{\s(\pm1,\pm1)}_\alpha]=0~,
\lb{Cd2}\\
&&
D^{\s(\pm4,0)}=D^{\s(\pm2,2)} D^{\s(\pm2,-2)}~,\qquad
D^{\s(0,\pm4)}=D^{\s(2,\pm2)} D^{\s(-2,\pm2)}~.\lb{Cd3}
\eea

Introduce the $r$-analytic  coordinates
\bea
&&x_r^\ab=x^\ab+{i\over2}[\theta_k^{\alpha\s(0,1)}
\theta^{\beta k\s(0,-1)}+ (\alpha\leftrightarrow\beta)]~,\lb{rx}\\
&& \theta_{\alpha}^{k\s(0,\pm1)}=v^{\s(0,\pm1)}_a\theta^{ka}_\alpha~.
\lb{rth}
\eea
These coordinates are natural for the $r$-analytic superfields
$\Phi_r(x^\ab_r,\theta_{\alpha}^{k\s(0,1)},v^{\s(0,\pm1)}_a)$ which
describe the alternative representations of the $D{=}3, \cN{=}8$
supersymmetry.

The spinor and harmonic derivatives in these coordinates have the
following form:
\bea
&&D^{k\s(0,1)}_\alpha=v^{\s(0,1)}_a D^{ka}_\alpha=
\partial^{k\s(0,1)}_\alpha~,\lb{rsp+}\\
&&D^{k\s(0,-1)}_\alpha=v^{\s(0,-1)}_a D^{ka}_\alpha=-
\partial^{k\s(0,-1)}_\alpha +
i\theta^{\beta k\s(0,-1)}\partial_\ab^r~,    \lb{rspin}\\
&&D_r^{\s(0,2)}=\partial_r^{\s(0,2)} - {i\over2}\theta^{\alpha\s(0,1)}_k
\theta^{k\beta\s(0,1)}
\partial_\ab^r + \theta^{\alpha\s(0,1)}_k
 \partial^{k\s(0,1)}_\alpha ~.\lb{rharm}
\eea

The constraints \p{C8} are equivalent to the $r$-analyticity condition
\be
D^{k\s(0,1)}_\alpha W^{\s(0,2)}(\zeta_r,v)=0 \lb{ranalit}
\ee
and the following $r$-harmonic conditions:
\be
 D_r^{\s(0,2)} W^{\s(0,2)}=0~.
\lb{C16b}
\ee

One can consider the component decomposition of this representation
of the $l$-vector multiplet
which is equivalent to {\it the $r$-linear analytic
multiplet}
\bea
&W^{\s(0,2)}(\zeta_r,v)=
v_a^{\s(0,1)}v_b^{\s(0,1)}\Phi^{ab}(x_r)+\theta^{\alpha\s(0,1)}_k
v_a^{\s(0,1)}\lambda^{ka}_\alpha(x_r)\nn&\\
&+\theta^{\alpha\s(0,1)}_k\theta^{\s(0,1)}_{l\alpha} X^{kl}(x_r)+
 \theta^{\alpha\s(0,1)}_k\theta^{\beta k\s(0,1)}[F_\ab + {i\over2}
\partial_\ab^r\Phi^{ab}v_a^{\s(0,1)}v_b^{\s(0,-1)}]+
O[(\theta^{\s(0,1)}_k)^3]~,\lb{analstr}&
\eea
where $F_\ab$ is the transversal field-strength of the Abelian gauge
field.

The constraints on $W^{\s(0,2)}$ are evident in the $BHS$
representation
\be
W^{\s(0,2)}=-iD^{\s(2,2)}V_l^{\s(-2,0)}=-i\int du D^{\s(-2,2)}
V_l^{\s(2,0)}~.\lb{C16c}
\ee
It is clear that this representation of the $l$-vector multiplet
is equivalent to the representation \p{C7}.

Consider the $r$-harmonic decomposition of the full spinor measure
\be
d^{\s8}\theta= D^{\s(0,-4)}D^{\s(0,4)}~.\lb{C17b}
\ee
Using this decomposition and Eqs.(\ref{C11},\ref{C14}) and \p{C16c} we
can construct an equivalent form of the effective action in  the
$r$-analytic superspace
\be
S_\3=\int d^\3x D^{\s(0,-4)} dv [W^{\s(0,2)}]^2 F_\3[W^{\s(0,2)},~
v^{\s(0,\pm1)}]~.\lb{C17}
\ee
It should be underlined that this action with the gauge-invariant
analytic Lagrangian can be generalized to the case of non-Abelian $SYM$
theory.

Let us consider now the set of $l$-analytic prepotentials
$V^{\s(2,0)}_{l\B}$ in the $[U(1)]^p$ gauge theory and the corresponding
$r$-analytic superfields $W^{\s(0,2)}_\B(V^{\s(2,0)}_{l\B})$. The
effective action of this theory in the $r$-analytic superspace is
\be
S_\3^p=\int d^\3x D^{\s(0,-4)} dv \sum\limits^p_{\B,\C=1} W^{\s(0,2)}_\B
W^{\s(0,2)}_\C F^\3_{\s BC}[W^{\s(0,2)}_\A,~v^{\s(0,\pm1)}]~,
\lb{C18b}
\ee
where $F_{\s BC}$ are real $q{=}(0,0)$ functions of the superfields
$W^{\s(0,2)}_1,\ldots W^{\s(0,2)}_p$ and $v$-harmonics. The corresponding
effective action in the full superspace contains the matrix 
superpotential of the $[U(1)]^p$ gauge theory
\bea
&&S_\3^p=\sum\limits^p_{\B,\C=1}\int d^{\3}\!x\,d^{\s8}\!\theta\,du\,dv\;
V^{\s(2,0)}_{l\B} V^{\s(-2,0)}_{l\C} f^\3_{\s BC}(W^{ab}_1,\ldots
W^{ab}_p)~,\lb{C18d}\\
&&f^\3_{\s BC}(W^{ab}_1,\ldots W^{ab}_p)=\int dv F^\3_{\s BC}
[W^{\s(0,2)}_\A,~v^{\s(0,\pm)}]~.
\lb{C18c}
\eea

The $v$-integral  representation of the matrix
superpotential satisfies the following conditions:
\bea
&&\frac{\partial}{\partial W^{cd}_\M}\frac{\partial}{\partial W_{cd\N}}
f^\3_{\s BC}\equiv\Delta^{\M\N}f^\3_{\s BC}=0~,\lb{C19}\\
&& D^{\s(+2)}_\ab f^\3_{\s BC}=0~.
\lb{C20}
\eea
This can be proved with the help of the harmonic decomposition of
the $SU_r(2)$-invariant operator $\Delta^{\M\N}$ by analogy with
\p{C15}
\be
\Delta^{\M\N}~\sim~\frac{\partial}{\partial W^{\s(0,2)}_\M}
\frac{\partial}{\partial W^{\s(0,-2)}_\N}-\frac{\partial}
{\partial W^{\s(0,0)}_\M}
\frac{\partial}{\partial W^{\s(0,0)}_\N}~.\lb{Lptyp}
\ee
These conditions guarantee the gauge invariance of $S_\3^p$ in the full
superspace.

The  $r$-forms of the hypermultiplet constraints
have been discussed in ref.\cite{Z3}. Consider the superfield
constraints for these hypermultiplets in the framework of $BHS$
\bea
&& D^{\s(\pm1,1)}_\alpha q_r^{a\s(0,1)}=0~,\qquad  q_r^{a\s(0,1)}=
(q_r^{\s(0,1)},\bar{q}_r^{\s(0,1)})~,\lb{C21}\\
&& D^{\s(\pm1,1)}_\alpha \omega_r=0~,\qquad D_l^{(\s\pm2,0)} (\omega_r,
q_r^{a\s(0,1)})=0~,
\lb{C22}
\eea
where $D_l^{\s(\pm2,0)}$ are the $l$-harmonic derivatives.

These hypermultiplets are dual to each other and also to the $r$-linear
analytic multiplet
\be
q_r^{a\s(0,1)}=v^{a\s(0,1)}\omega_r +
v^{a\s(0,-1)} L^{\s(0,2)}~.\lb{C23}
\ee
The duality relation between the $\omega_r$ and $r$-linear  multiplet
is described by the action
\be
\int d^\3x D^{\s(0,-4)} dv\{ \omega_r[D_r^{\s(0,2)}L^{\s(0,2)}]  +
 F^{\s(0,4)}[L^{\s(0,2)},~v^{\s(0,\pm1)}]\}~,\lb{C24}
\ee
where $F^{\s(0,4)}$ is an arbitrary $r$-analytic function.

It is clear that the $l$-analytic hypermultiplets $q_l^{(1,0)}$ and
$\omega_l$ are dual to the alternative $r$-version of the vector
multiplet which can be described by the $r$-analytic prepotential
$V_r^{\s(0,2)}$.

Thus, the $l$-analyticity allows us to solve the constraints of
the $l$-vector multiplet and $l$-hypermultiplets, while the
$r$-analyticity generalizes the holomorphicity and chirality in the
$HS$ description of low-energy gauge actions and duality symmetries.

\setcounter{equation}0
\section{\lb{D} Two-dimensional (4,4) harmonic superspaces}

The $D{=}2,$ (4,4)-supersymmetric field theories describe 1-branes
probing a background with 5-branes in M-theory \cite{CHS,DS,Wi}.
 The two-dimensional (4,4) and (4,0) $\sigma$-models have been 
discussed in the field-component formalism  and in the framework of
the ordinary or harmonic superspaces \cite{GHR}-\cite{GS}.
The $2D$ mirror symmetry and the (4,4)
gauge theory has been considered in the component formalism and in the
(2,2) superspace \cite{MP,DS,Wi}. We shall  study the geometry
of this theory in the manifestly covariant harmonic formalism which
is convenient for the superfield quantum calculations. Three types
of Grassmann analyticities will be used to classify the (4,4)
supermultiplets, their interactions and $2D$ duality symmetries.

The maximum automorphism group of the (4,4) superspace is $SO_l(4){\times}
SO_r(4)$; however, we shall mainly use the group $R_\2{=}SU_c(2){\times}
SU_l(2){\times}SU_r(2)$. Let us choose the left and right coordinates in
the (4,4) superspace
\be
z_l=(y,~\theta^{i\alpha})~,\qquad\qquad z_r=
(\bar{y},~\bar{\theta}^{ia})~,\lb{D1}
\ee
where $y=(1/\sqrt{2})(t+x)~$ and $\bar{y}=(1/\sqrt{2})(t-x)$ are the
light-cone $2D$ coordinates; and the following types of 2-spinor indices
are used: $  i, k, \ldots$ for $SU_c(2)$; $~\alpha,\beta\ldots$ for
$SU_l(2)$ and $~a, b\ldots$ for $SU_r(2)$, respectively. The $SO(1,1)$
weights of coordinates are $(1,~1/2)$ for $z_l$ and $(-1,~-1/2)$ for
$z_r$. The algebra of spinor derivatives in this superspace
\bea
&& \{D_{k\alpha},D_{l\beta}\}=i\varepsilon_{kl}\varepsilon_\ab\partial_y 
~,\lb{D2}\\
&& \{\bar{D}_{ka},\bar{D}_{lb}\}=i\varepsilon_{kl}\varepsilon_{ab}
\bar{\partial}_y ~,\lb{D3}\\
&& \{D_{k\alpha},\bar{D}_{lb}\}=i\varepsilon_{kl}Z_{\alpha b}\lb{D4}
\eea
contains the central charges $Z_{\alpha b}$.

The $CB$-geometry of the (4,4) $SYM$ theory is described by the
superfield constraints
\bea
&& \{\nabla_{k\alpha},\nabla_{l\beta}\}=i\varepsilon_{kl}\varepsilon_\ab
   \nabla_y~,\lb{D5}\\
&& \{\bar{\nabla}_{ka},\bar{\nabla}_{lb}\}=i\varepsilon_{kl}
\varepsilon_{ab} \bar{\nabla}_y~,\lb{D6}\\
&& \{\nabla_{k\alpha},\bar{\nabla}_{lb}\}= i\varepsilon_{kl}
W_{\alpha b}~, \lb{D7}
\eea
where $\nabla_\M{=}D_\M+A_\M$ is the covariant derivative for the
corresponding coordinate. The gauge-covariant superfield $W_{\alpha b}$
satisfies the constraints of the (4,4) vector multiplet which are
equivalent to the constraints of the so-called twisted multiplet
\cite{GHR,GoI}.

The authors of refs.\cite{IS1,IS2,Iv1} have discussed  three types  of
harmonics: $u^\pm_i{=}u^{\s(\pm1,0,0)}_i$ for $SU_c(2)/U_c(1)$ ; $
l^{\s(0,\pm1,0)}_\alpha$ for $SU_l(2)/U_l(1)$; and $r^{\s(0,0,\pm1)}_a$
for $SU_r(2)/U_r(1)$.  We use the notation with 3 charges in the
triharmonic superspace $(THS)$ and the standard notation in the
$c$-harmonic superspace $HS_c$. The basic geometric structures of the
gauge theory are mainly connected with the $c$-harmonics $u^\pm_i$ and 
the corresponding analytic coordinates
\be
\zeta_c=(y_c,~\theta^{\alpha+}),
~\theta^{\alpha-}~,\qquad \bar{\zeta}_c=(\bar{y}_c,~
\bar{\theta}^{a+}),~\bar{\theta}^{a-}.\lb{canal}
\ee

 The $HS_c$ spinor derivatives  and harmonic derivatives
have the following form in the case of vanishing central charges:
\bea
&&D^+_\alpha=\partial^+_\alpha~,\qquad D^-_\beta=-\partial^-_\alpha-
i\theta^-_\alpha\partial_y^c~,\lb{D10}\\
&&\bar{D}^+_a=\partial^+_a~,\qquad \bar{D}^-_a=-\bar{\partial}^-_a-
i\bar{\theta}^-_a\bar{\partial}_y^c~,\lb{D10b}\\
&&\Dp_c=\dpp_c + {i\over2}\theta^{+\alpha}\theta^+_\alpha \partial_y^c +
{i\over2}\bar{\theta}^{a+}\bar{\theta}^+_a\bar{\partial}_y^c+
\theta^{\alpha+} \partial^+_\alpha +\bar{\theta}^{a+} \partial^+_a~.
\lb{D10c}
\eea

The basic combinations of the spinor derivatives are
\be
(D^\pm)^2={1\over2}D^{\pm\alpha}D^\pm_\alpha~,\quad (\bar{D}^\pm)^2=
{1\over2}\bar{D}^{a\pm}
\bar{D}^\pm_a~,\quad (D^\pm)^4=(D^\pm)^2(\bar{D}^\pm)^2~.\lb{D11}
\ee

The $c$-harmonic projections of the constraints (\ref{D5}-\ref{D7})
are equivalent to the integrability conditions of the $c$-analyticity
by analogy with the $D{\geq}3,~\cN{=}8$ theories
\be
\{\nabla_\alpha^+,\nabla_\beta^+\}=\{\nabla_\alpha^+,\bar{\nabla}_b^+\}=
\{\bar{\nabla}_a^+,\bar{\nabla}_b^+\}=0~, \lb{D11b}
\ee
where $\nabla_\alpha^+=u^+_i\nabla_\alpha^i$ and $\bar{\nabla}_a^+=u^+_i
\bar{\nabla}_a^i$.

The prepotential of the (4,4) gauge theory in  $HS_c$ is the $c$-analytic
harmonic connection $\Vp_c(\zeta_c~,\bar{\zeta}_c,~u)\equiv
V_c^{\s(2,0,0)}$ which determines the second harmonic connection
$\Vm_c\equiv V_c^{\s(-2,0,0)}$.
The $WZ$ gauge for this prepotential has the following form:
\bea
&(\Vp_c)_{\s WZ}=\theta^{\alpha+}\bar{\theta}^{b+}\Phi_{\alpha b}
(y_c\,,\bar{y}_c)
+(\theta^+)^2 \bar{A}(y_c\,,\bar{y}_c)+(\bar{\theta}^+)^2 A(y_c\,,
\bar{y}_c)&\nn\\
&+(\bar{\theta}^+)^2\theta^{\alpha+} u^-_i\lambda_\alpha^i(y_c\,,
\bar{y}_c)+(\theta^+)^2\bar{\theta}^{+a} u^-_i\bar{\lambda}_a^i(y_c\,,
\bar{y}_c)+i(\theta^+)^2(\bar{\theta}^+)^2 u^-_k u^-_j X^{kj}(y_c\,,
\bar{y}_c) & \lb{2WZ}
\eea
where the components of the $2D$ vector multiplet are defined.

The gauge-covariant Abelian superfield strength can be constructed by
analogy with $D{=}3$
\be
W_{\alpha b}\equiv (\sigma^m)_{\alpha b}W_m=-iD^+_\alpha \bar{D}^+_b
\Vm_c \lb{D12}
\ee
where $(\sigma^m)_{\alpha b}$ are the  Weyl matrices for $SU_l(2){\times}
SU_r(2)$ and $W_m$ is the 4-vector representation of this superfield.

The Abelian superfield $W_{\alpha b}$ does not depend on harmonics, and
the constraints for this superfield are
\be
 D^+_\alpha W_{\beta b}={1\over2}\varepsilon_\ab D^{+\rho}W_{\rho b}~,
\qquad \bar{D}^+_a W_{\alpha b}={1\over2}\varepsilon_{ab}\bar{D}^{+c}
W_{\alpha c}~.\lb{D13}
\ee
We shall use also the following consequences of these relations:
\be
 (D^+)^2 W_{\alpha a}=
(\bar{D}^+)^2 W_{\alpha a}=0~.\lb{D14}
\ee

The $c$-analytic (4,4) hypermultiplets $q^+_c$ and $\omega_c$ have
the minimal interactions with $\Vp_c$. The corresponding $HS$ Feynmann
rules can be formulated by analogy with ref. \cite{GIOS2}. The  $HS$
perturbative methods can be useful in the analysis of the
vector-hypermultiplet Matrix models with (8,8) supersymmetry, however,
we shall discuss here the general symmetry framework for such
calculations.

The universal harmonic construction of the $U(1)$ effective action
with 8 supercharges has the following form in  the case $D{=}2$:
\be
S_\2=\int d^{\2}x d^{\s8}\theta du \Vp_c \Vm_c f_\2(W_m)~,\lb{D15b}
\ee
where $d^{\2}x=dt dx\equiv dy d\bar{y}$.
The gauge invariance imposes the following constraints:
\be
(D^+)^2 f_\2(W_m)=(\bar{D}^+)^2 f_\2(W_m)=0~.
\lb{D15}
\ee

Using the Eqs.\p{D14} one can prove that
the  (4,4) superpotential satisfies the $4D$ Laplace equation
\be
\Delta_\4^w f_\2(W_m)=0~,\qquad \Delta_\4^w=\Bigl(\frac{\partial}
{\partial W_m}\Bigr)^2~.
\lb{D16}
\ee
The analogous $4D$ Laplace equation in the (4,4) $\sigma$-models has
been discussed, for instance, in refs.\cite{CHS,GS}.

The $R_\2$-invariant solution of this equation determines
uniquely the exact      superpotential of the (4,4) gauge theory
\be
f^\R_\2(w_\2)= g^{-2}_\2+ k_\2 w_\2^{-2}~,\quad w_\2=
\sqrt{W^{\alpha a}W_{\alpha a}}~.\lb{D17}
\ee
The same function of the (2,2) superfields generates the
$R_\2$-invariant  K\"{a}hler potential of the $D{=}2, (4,4)$ gauge
theory \cite{DS}. Note that the K\"{a}hler potential of the (2,2)
formalism is gauge-invariant by definition, and the  $4D$ Laplace
equation arises in this approach from the restrictions of the (4,4)
supersymmetry; while in our  formulation  the analogous condition on
the (4,4) superpotential \p{D16} follows from the gauge invariance.
The manifestly (4,4) covariant formalism of the harmonic gauge theory
simplifies the proof of the non-renormalization theorem.

The  $THS$ projections of the $2D$ spinor derivatives are
\be
D^{\s(\pm1,\pm1,0)}=u^{\s(\pm1,0,0)}_i l^{\s(0,\pm1,0)}_\alpha
D^{\alpha i}~,\qquad\bar{D}^{\s(\pm1,0,\pm1)}=u^{\s(\pm1,0,0)}_i
r^{\s(0,0,\pm1)}_a\bar{D}^{ai}~.\lb{D18}
\ee
The $rl$-version of the $c$-vector multiplet \p{D12} has the following
form:
\be
W^{\s(0,1,1)}=-iD^{\s(1,1,0)}\bar{D}^{\s(1,0,1)}V_c^{\s(-2,0,0)}=-i
\int du D^{\s(-1,1,0)}\bar{D}^{\s(-1,0,1)}V_c^{\s(2,0,0)}~.\lb{D19}
\ee

By construction, this superfield satisfies the conditions of the
$rl$-analyticity
\be
D^{\s(\pm1,1,0)} W^{\s(0,1,1)}=0~,\qquad \bar{D}^{\s(\pm1,0,1)}
W^{\s(0,1,1)}=0 \lb{D20}
\ee
and the harmonic conditions
\be
D_c^{\s(\pm2,0,0)} W^{\s(0,1,1)}=D_l^{\s(0,2,0)}W^{\s(0,1,1)}=
D_r^{\s(0,0,2)}W^{\s(0,1,1)}=0~. \lb{D21}
\ee

The analogous  constraints on the $rl$-harmonic  superfield
$q^{\s(1,1)}$ have been considered
in ref.\cite{IS1} (this notation does not indicate the $U_c(1)$ charge).
 Note that the vector multiplet \p{D19} contains the field-strength of
the $2D$ vector field instead of the auxiliary scalar component in the
superfield $q^{\s(1,1)}$.

The $c$-analyticity become manifest in the coordinates \p{canal}.
Let us consider the analogous  $rl$-analytic coordinates which
help to solve the conditions \p{D20}
\bea
&&\zeta_l=(y_l, \theta^{\s(\pm1,1,0)})~,\qquad\theta^{\s(\pm1,\pm1,0)}=
u^{\s(\pm1,0,0)}_i l^{\s(0,\pm1,0)}_\alpha\theta^{i\alpha}~,
\lb{D21b}\\
&&y_l=y+{i\over2}[\theta^{\s(1,-1,0)}\theta^{\s(-1,1,0)}-
\theta^{\s(-1,-1,0)}\theta^{\s(1,1,0)}]~,\lb{D21d}\\
&&\bar{\zeta}_r=(\bar{y}_r, \bar{\theta}^{\s(\pm1,0,1)})~,\qquad
\bar{\theta}^{\s(\pm1,0,\pm1)}=u^{\s(\pm1,0,0)}_i r^{\s(0,0,\pm1)}_a
\bar{\theta}^{ia}~,\\
\lb{D21c}
&&y_r=y+{i\over2}[\bar{\theta}^{\s(1,0,-1)}\bar{\theta}^{\s(-1,0,1)}-
\bar{\theta}^{\s(-1,0,-1)}\bar{\theta}^{\s(1,0,1)}]~.\lb{D21e}
\eea

The spinor and harmonic derivatives have the following form in these
coordinates:
\bea
&& D^{\s(\pm1,1,0)}=\pm\partial^{\s(\pm1,1,0)}~,\qquad
D^{\s(\pm1,-1,0)}=\mp\partial^{\s(\pm1,-1,0)}+i\theta^{\s(\pm1,-1,0)}
\partial^l_y~,\lb{D22b}\\
&& \bar{D}^{\s(\pm1,0,1)}=\pm\bar{\partial}^{\s(\pm1,0,1)}~,\qquad
\bar{D}^{\s(\pm1,0,-1)}=\mp\bar{\partial}^{\s(\pm1,0,-1)}+
i\bar{\theta}^{\s(\pm1,-1,0)}
\bar{\partial}^r_y~,\lb{D22c}\\
&& D_l^{\s(0,2,0)}=\partial_l^{\s(0,2,0)}+i\theta^{\s(1,1,0)}
\theta^{\s(-1,1,0)}\partial^l_y +\theta^{\s(1,1,0)}\partial^{\s(-1,1,0)}
+ \theta^{\s(-1,1,0)}\partial^{\s(1,1,0)}~,\lb{D22d}\\
&& D_r^{\s(0,0,2)}=\partial_r^{\s(0,0,2)}+i\bar{\theta}^{\s(1,0,1)}
\bar{\theta}^{\s(-1,0,1)}\bar{\partial}^r_y +\bar{\theta}^{\s(1,0,1)}
\bar{\partial}^{\s(-1,0,1)} +
\bar{\theta}^{\s(-1,0,1)}\bar{\partial}^{\s(1,0,1)}~.\lb{D22e}
\eea

The $c$-analytic coordinates in the $THS$ notation are
\bea
&&\zeta_c=(y_c,\theta^{\s(1,\pm1,0)})~,\qquad
y_c=y + {i\over2}[\theta^{\s(-1,1,0)}  \theta^{\s(1,-1,0)}+
\theta^{\s(1,1,0)}  \theta^{\s(-1,-1,0)}]  ~,
\lb{D8m}\\
&&\bar{\zeta}_c=(\bar{y}_c,\bar{\theta}^{\s(1,0,\pm1)})~,\qquad
\bar{y}_c=\bar{y} + {i\over2}[\bar{\theta}^{\s(-1,0,1)}
\bar{\theta}^{\s(1,0,-1)} +\bar{\theta}^{\s(1,0,1)}
\bar{\theta}^{\s(-1,0,-1)}]  ~.
\lb{D9m}
\eea
It is important that all  coordinates $\zeta_c,\bar{\zeta}_c,\zeta_l$
and $\bar{\zeta}_r$ are  separately real with respect to the
corresponding conjugation. Of course, all these sets of coordinates
are irreducible with respect to the supersymmetry transformations.

The solution of the $4D$ Laplace equation \p{D16} has the
simple harmonic representation
\be
f_\2(W_{\alpha a})=\int dl dr F_\2[W^{\s(0,1,1)},~l,~r]~,
\lb{D22}
\ee
where $F_\2$ is the real function and $W^{\s(0,1,1)}=l_\alpha^{\s(0,1,0)}
r_a^{\s(0,0,1)}W^{\alpha a}$ \p{D19}. The proof is based on the $THS$
decomposition of the $4D$ Laplace operator
\be
\frac{\partial}{\partial  W^{\alpha b}}
\frac{\partial}{\partial W_{\alpha b}}
\sim {\partial\over \partial W^{\s(0,1,1)}}
{\partial\over \partial W^{\s(0,-1,-1)}}-{\partial\over \partial
W^{\s(0,1,-1)}}{\partial\over \partial W^{\s(0,-1,1)}}~.\lb{D23}
\ee
Note that the formal change of the density in \p{D22}
\be
F_\2[W^{\s(0,1,1)},~l,~r]~\rightarrow~F^\prime[W^{\s(0,1,1)},
W^{\s(0,1,-1)},~l,~r] \lb{D23b}
\ee
does not produce more general superpotentials. This can be easily shown
for the polynomial solutions of Eq.\p{D16}.

Consider the $THS$ decomposition of the Grassmann measure
\bea
&& d^{\s8}\theta=D^{\s(0,-2,-2)}D^{\s(0,2,2)}\lb{D27c}~,\\
&&D^{\s(0,\pm2,\pm2)}=D^{\s(1,\pm1,0)}D^{\s(-1,\pm1,0)}
\bar{D}^{\s(1,0,\pm1)}\bar{D}^{\s(-1,0,\pm1)}~.\lb{D27b}
\eea
Using this decomposition and Eqs.(\ref{D15},\ref{D19}) we can obtain the
following equivalent representation of the effective (4,4) action in the
$rl$-analytic superspace:
\be
S_\2=\int dl dr d^{\2}x D^{\s(0,-2,-2)}
[W^{\s(0,1,1)}]^2 F_\2[W^{\s(0,1,1)},~l,~r]~. \lb{D27}
\ee

One can construct the effective (4,4) action for the gauge group
$[U(1)]^p$ in the $rl$-analytic and full superspaces by analogy with the
case $D{=}3$ (\ref{C18b},\ref{C18d}).

An analogous action of the $q^{\s(1,1)}$ multiplet and dual superfields
$\omega^{\s(\pm1,\mp1)}$ has been considered in refs.\cite{IS1,IS2,Iv1}.
The relation between the $c$-analytic gauge superfield and $rl$-analytic
hypermultiplets is a specific manifestation of the $2D$ mirror symmetry
\cite{MP}. Consider the $rl$-analytic superfield $Q^{\s(0,1,1)}$ in our
notation. The $r$- and $l$-harmonic constraints \p{D21} can be introduced
via the $rl$-analytic Lagrange multipliers
\bea
&&S(Q,\omega)=\int dl dr d^{\2}x D^{\s(0,-2,-2)}[F^{\s(0,2,2)}
(Q^{\s(0,1,1)},~r,~l)\nn\\
&&+\omega^{\s(0,1,-1)}D_l^{\s(0,2,0)}Q^{\s(0,1,1)}+
\omega^{\s(0,-1,1)}D_r^{\s(0,0,2)}Q^{\s(0,1,1)}]~.\lb{dual2}
\eea

The triharmonic superspace is convenient for the classification
of the (4,4) supermultiplets. Let us consider, for instance, the
$cr$-analytic superfield $Q_{cr}^{\s(1,0,1)}(\zeta_c,\bar{\zeta}_r,u,r)$
satisfying the subsidiary harmonic conditions
\be
D_c^{\s(2,0,0)}Q_{cr}^{\s(1,0,1)}=0~,\qquad
D_r^{\s(0,0,2)}Q_{cr}^{\s(1,0,1)}=0~,\lb{D28}
\ee
where the analytic coordinates \p{D8m} and \p{D21c} are used.
The $cl$-analytic superfield $Q_{cl}^{\s(1,1,0)}(\bar{\zeta}_c,
\zeta_l,u,l)$ can be defined analogously.

Thus, the alternative $HS$ structures and their
embedding to the general triharmonic superspace
are natural for the off-shell geometric description of the (4,4)
supersymmetric theories.

\setcounter{equation}0
\section{\lb{E} One-dimensional harmonic superspaces}

The one-dimensional $\sigma$-models have been considered in the component
formalism and also in the framework of the superspaces with $\cN{=}1, 2$
and 4 \cite{BP,ISm,GPS}. Recently, the $\cN{=}4$ superspace has been used
 for the proof of the non-renormalization theorem in the $\cN{=}8$
gauge theory \cite{DE}.  This quantum-mechanical model describes
$D0$-probes moving in different $D4$-brane backgrounds.

It is interesting to study these models in the framework of the 
manifestly supersymmetric harmonic approach.
We shall consider the $D{=}1,~\cN{=}8$ superspace which is based on the
maximum automorphism group $R_{\s1}{=}SU_c(2){\times} Spin(5)$ and has
coordinates $z{=}(t,~\theta^\dal_i) $ ( $i,k,l\ldots$ are the 2-spinor
indices and $\alpha,\beta,\rho\ldots$ are the 4-spinor indices of the
group $Spin(5){=}USp(4)$). The algebra of spinor derivatives is
\be
\{D^k_\dal,D^l_\drh\}=i\varepsilon^{kl}\Omega_\diar\partial_t
+i\varepsilon^{kl}Z_\diar~,\lb{E1}
\ee
where $Z_\diar$ are central charges and $\Omega_\diar$ is the 
antisymmetric $Spin(5)$ metric.

Conjugation rules in the group $Spin(5)$ differ from the corresponding
rules in $Spin(4,1)$ \p{B5c}
\be
\overline{\theta^\dal_i}=\theta^i_\dal~,\quad \overline{\Omega_\diar}=
-\Omega^\diar~,\quad \overline{Z_\diar}=Z^\diar~.
\lb{E1b}
\ee

The $CB$-geometric superfield constraints of the $\cN{=}8~SYM$ theory are
\be
\{\nabla^k_\dal,\nabla^l_\drh\}=i\varepsilon^{kl}\Omega_\diar(\partial_t+
A_t)+i\varepsilon^{kl}W_\diar~,\lb{E2}
\ee
where a traceless bispinor (or 5-vector) superfield representation of
the $1D$ vector multiplet $W_\diar(z)$ is defined.

The harmonics $u^\pm_i$ can be used for a construction of the $D{=}1$
$c$-analytic coordinates $\zeta_c{=}(t_c,~\theta^{+\dal})$
\be
t_c=t+{i\over2}\theta^\dal_k\theta_{l\dal}u^{k+}u^{l-}~,\qquad
\theta^{+\dal}=u^+_k\theta^{k\dal}~.\lb{E3}
\ee

The algebra of the $c$-harmonized $1D$ spinor derivatives resembles
the corresponding algebra of the $5D$ derivatives (\ref{B7}-\ref{B9})
with $Spin(5)$ indices instead of the $Spin(4,1)$ indices. In the case of
vanishing central charges we have
\bea
&& \Dalp=\partial^+_\dal~,\qquad \Dalm=-\partial^-_\dal +i\theta^-_\dal
\partial_t^c~,\lb{E4}\\
&& \Dp_c=\dpp_c - {i\over2}\theta^{\dal+} \theta^+_\dal\partial_t^c
+\tdap \partial^+_\dal ~.\lb{E4b}
\eea

The constraints \p{E2} correspond to the integrability conditions of
the $c$-analyticity
\be
\{\nabla^+_\dal, \nabla^+_\dga\}=0~,\qquad \nabla^+_\dal=u^+_i
\nabla^i_\dal~.\lb{E4c}
\ee
The $c$-analytic prepotential $\Vp_c(\zeta_c,u)$ describes the $1D$
vector multiplet (or 8+8 $\sigma$-model) and contains  the pure gauge
one-dimensional field
$A$
\bea
&&(\Vp_c)_{\s WZ}=\Theta^{(+2)}A(t_c)+\Theta^{(+2)\diar} \Phi_\diar(t_c)
\nn\\
&&+\Theta^{(+2)}\theta^{\dal+}u^-_k\lambda^k_\dal(t_c)+i[\Theta^{(+2)}]^2
u^-_k u^-_j X^{kj}(t_c)~,\lb{1WZ}
\eea
where the notation \p{B9c} is used. Of course, one can use the subsidiary
gauge condition $A(t_c){=}0$.

 The  basic superfield  in the $D{=}1,~
\cN{=}8$ Abelian gauge theory has the following form:
\be
W_\diar\equiv {1\over2}(\Gamma^\bm)_\diar W_\bm=-iD^{(+2)}_\diar\Vm                                  ~,
\qquad \Omega^\diar W_\diar=0~,\lb{E5}
\ee
where the  $\Gamma$  matrices of $Spin(5)$ are introduced.

The constraints for this superfield are satisfied by construction
\bea
&& \Dalp W_\dibg={2\over5}\Omega_\diab D^{+\dsi}W_\disg
-{2\over5}\Omega_\diag D^{+\dsi}W_\disb +{1\over5}\Omega_\dibg
D^{+\dsi}W_\disa ~,\lb{E6}\\
&& D^{(+2)} W_\diar=0~,\qquad D^{\s\pm\pm} W_\diar=0~.\lb{E7}
\eea

These constraints are equivalent to the conditions of different (twisted)
chiralities for the superfields $W_{13},W_{14},W_{23}$ and
$W_{24}$ , e.g.
\be
D^\pm_1 W_{13}=D^\pm_3 W_{13}=0~,\qquad D^\pm_1 W_{14}=D^\pm_4 W_{14}=0~.
\lb{E8}
\ee

The $c$-analytic hypermultiplets $q^+(\zeta_c,u)$ and $\omega_c
(\zeta_c,u)$ can be introduced by  analogy with $HS$ of higher
dimensions. These superfields have the $R_{\s1}$-invariant minimal
interactions with the prepotential $\Vp_c$. We do not consider here
the $HS$ perturbative analysis of this model which can describe the
$D0$-$D4$ interactions in Matrix theory and restrict ourselves to
the study of a general symmetry framework for these calculations.

The $D{=}1$ low-energy $U(1)$-gauge action has the following universal
form:
\be
S_{\s1}=\int dt d^{\s8}\theta du\; \Vp_c \Vm_c\;f_{\s1}(W_\bm)
~.\lb{E9}
\ee

Using the constraint \p{E6} one can prove that the gauge invariance of
$S_{\s1}$ is equivalent to the $5D$ Laplace equation for the
superpotential
\be
D^{(+2)}\,f_{\s1}(W_\bm) =0~\rightarrow~ \Delta_\5^w f_{\s1}(W_\bm) =0~.
\lb{E10}
\ee

The $R_{\s1}$-invariant $D{=}1$ superpotential
\be
f^\R_{\s1}(w_{\s1})= g^{-2}_{\s1}+ k_{\s1} w_{\s1}^{-3}~,
\qquad w_{\s1}=( W^\dirs W_\dirs)^{1/2} \lb{E11}
\ee
is the unique solution of this equation. The non-renormalizability
of this superpotential is protected by the $Spin(5)$-invariance and
the $\cN{=}8$ supersymmetry.
Note that the same function determines the K\"{a}hler potential of the
$D{=}1$ gauge theory in the ${\cal N}{=}4$ superfield formalism
\cite{DE}.

By analogy with the cases $D{=}2$ and 3, the geometric description of
the $D{=}1,~\cN{=}8$ models requires the use of harmonic variables
for the whole group $SU(2){\times}Spin(5)$.
Let us introduce now the biharmonic $1D$-superspace using the
$SU_c(2)$ harmonics $u^{\s(\pm1,0,0)}_i{=}u^\pm_i$  and harmonics
$v^{\s(0,\pm1,0)}_\dal,~v^{\s(0,0,\pm1)}_\dal$ of the group $USp(4)$
\cite{IKNO}. The basic relations for the $v$-harmonics are
\bea
&& \Omega^\diar v^{\s(0,a,0)}_\dal v^{\s(0,-b,0)}_\drh=\delta^{ab}~,
\lb{E12}\\
&& \Omega^\diar v^{\s(0,0,a)}_\dal v^{\s(0,0,-b)}_\drh=\delta^{ab}~,
\lb{E13}~,\\
&& \Omega^\diar v^{\s(0,a,0)}_\dal v^{\s(0,0,b)}_\drh=0~.\lb{E14}
 \eea
where $a,b=\pm1$ and $\delta_{ab}$ is the Kronecker symbol. These
harmonics determine the 8-dimensional coset space $H_8{=}USp(4)/
U(1){\times}U(1)$.

The harmonic derivatives $D_v^{\s(0,\pm2,0)},~D_v^{\s(0,0,\pm2)}$ and
$D_v^{\s(0,\pm1,\pm1)}$ are defined in ref.\cite{IKNO}
\bea
&& D_v^{\s(0,\pm2,0)}v^{\s(0,\mp1,0)}_\dal=v^{\s(0,\pm1,0)}_\dal~,
\qquad D_v^{\s(0,0,\pm2)}v^{\s(0,0,\mp1)}_\dal=v^{\s(0,0,\pm1)}_\dal~,
 \lb{E14b}\\
&&D_v^{\s(0,\pm1,\pm1)}v^{\s(0,0,\mp1)}_\dal=v^{\s(0,\pm1,0)}_\dal~,
\qquad
D_v^{\s(0,\pm1,\pm1)}v^{\s(0,\mp1),0}_\dal=v^{\s(0,0,\pm1)}_\dal~,
\lb{E14e}\\
&& D_v^{\s(0,\pm2,0)}v^{\s(0,\pm1,0)}_\dal=D_v^{\s(0,\pm2,0)}
v^{\s(0,0,\pm1)}_\dal=D_v^{\s(0,0,\pm2)}v^{\s(0,\pm1,0)}_\dal=
D_v^{\s(0,0,\pm2)}v^{\s(0,0,\pm1)}_\dal=0~,\lb{E14c}\\
&& D_v^{\s(0,\pm1,\pm1)}v^{\s(0,0,\pm1)}_\dal=D_v^{\s(0,\pm1,\pm1)}
v^{\s(0,\pm1,0)}_\dal=0~.\lb{E14d}
\eea

 The  algebra of harmonic derivatives on $H_8$ contains also the
$U(1)$-charges $D^0_{v2}$ and $D^0_{v3}$. The harmonic derivatives on 
the coset $SU_c(2)/U_c(1)$ are $D_c^{\s(\pm2,0,0)}$ and $D^0_c$.

The $v$-harmonic representation of the general $1D$ superpotential
\p{E10} is
\bea
&&f_{\s1}(W_\diar)=\int dv F_{\s1}[W^{\s(0,1,1)},~v_\dal ]~,\lb{E20}\\
&&W^{\s(0,1,1)}=v^{\s(0,1,0)}_\dal v^{\s(0,0,1)}_\drh W^\diar~.
\lb{vproj}
\eea
where the real function $F_{\s1}$ of the single harmonic projection
$W^{\s(0,1,1)}$ and all components of the $v$-harmonics determines the
general solution of the $5D$ Laplace equation. The proof is based on the
$v$-harmonic decomposition of the operator $\Delta_\5^w$
using the $v$-harmonic completeness relation
\be
\Omega_\diar=v^{\s(0,-1,0)}_\dal v^{\s(0,1,0)}_\drh +
v^{\s(0,0,-1)}_\dal v^{\s(0,0,1)}_\drh -(\alpha\leftrightarrow\rho)~.
\lb{vcomplit}
\ee

Partial solutions can contain   the restricted density functions
$F_{\s1}$ of some harmonic components $v_{\s1}\ldots$ and correspond ,
for instance, to
the holomorphic functions of $W_{13}$ and/or $W_{14}$.

The $BHS$ spinor derivatives are
\be
D^{\s(\pm1,\pm1,0)}=u^{\s(\pm1,0,0)}_i v^{\s(0,\pm1,0)}_\dal D^{i\dal}~,
\qquad D^{\s(\pm1,0,\pm1)}=u^{\s(\pm1,0,0)}_i v^{\s(0,0,\pm1)}_\dal
D^{i\dal}~.\lb{E15}
\ee

The $v$-projection \p{vproj} of the basic gauge superfield  \p{E5}
can be written in terms of the   $c$-harmonic connections
\be
W^{\s(0,1,1)}=-iD^{\s(1,1,0)}D^{\s(1,0,1)}V_c^{\s(-2,0,0)}=
-i\int du D^{\s(-1,1,0)}D^{\s(-1,0,1)}V_c^{\s(2,0,0)}~.\lb{E16}
\ee
By construction, this superfield is $v$-analytic
\be
D^{\s(\pm1,1,0)}W^{\s(0,1,1)}=0,~\qquad D^{\s(\pm1,0,1)}W^{\s(0,1,1)}=0
\lb{E17}
\ee
and also satisfies the harmonic constraints
\be
D_c^{\s(\pm2,0,0)}W^{\s(0,1,1)}=0,~\quad \cD^\A_v W^{\s(0,1,1)}=0~,
\lb{E18}
\ee
where $\cD^\A_v$ is the triplet of harmonic derivatives conserving
the $v$-analyticity \p{E17}
\bea
&&\cD^\A_v=(D_v^{\s(0,1,1)},D_v^{\s(0,2,0)}, D_v^{\s(0,0,2)})~,
\lb{E19}\\
&&[\cD_\A,D^{\s(\pm1,1,0)}]=[\cD_\A,D^{\s(\pm1,0,1)}]=0~.\lb{E19b}
\eea

The $v$-analytic coordinates $\zeta_v{=}(t_v\,,\theta^{\s(\pm1,1,0)},
\theta^{\s(\pm1,0,1)})$ can be defined by analogy with \p{E3}
\bea
& t_v=t+{i\over2}[\theta^{\s(-1,-1,0)}\theta^{\s(1,1,0)}  -
\theta^{\s(1,-1,0)}\theta^{\s(-1,1,0)}-\theta^{\s(1,0,-1)}
\theta^{\s(-1,0,1)}+\theta^{\s(-1,0,-1)}\theta^{\s(1,0,1)}]~,&
\lb{E20b}\\
&\theta^{\s(\pm1,\pm1,0)}=u_i^{\s(\pm1,0,0)}v_\dal^{\s(0,\pm1,0)}
\theta^{i\dal}~,
\qquad \theta^{\s(\pm1,0,\pm1)}=u_i^{\s(\pm1,0,0)}v_\dal^{\s(0,0,\pm1)}
\theta^{i\dal}~.&
\lb{E20c}
\eea
These coordinates are convenient for the $v$-analytic superfields.

Using Eqs.(\ref{E16},\ref{E20}) we can obtain the
$v$-analytic
representation of the $1D$ effective action \p{E9}
\be
S_{\s1}=\int dv dt_v D^{\s(0,-2,-2)}
[W^{\s(0,1,1)}(\zeta_v)]^2  F_{\s1}[W^{\s(0,1,1)}(\zeta_v),~v_\dal ]~,
\lb{E21}
\ee
where the corresponding Grassmann measure is
\be
D^{\s(0,-2,-2)}=D^{\s(1,-1,0)}D^{\s(-1,-1,0)}D^{\s(1,0,-1)}
D^{\s(-1,0,-1)}~.\lb{E21b}
\ee

The invariant effective action for an arbitrary gauge group can be
constructed immediately in the $v$-analytic superspace. The
$v$-integral representation of the matrix superpotential for the gauge
group $[U(1)]^p$ in the full superspace satisfies the following
conditions:
\be
\frac{\partial}{\partial W^\digs_\M}\frac{\partial}{\partial W_{\digs\N}}
f^{\s1}_{\s BC}(W^\diar_1,\ldots W^\diar_p)=0~,\qquad
D^{(+2)}\,f^{\s1}_{\s BC}=0~.\lb{E21c}
\ee
The proof is analogous to the proof of the relations \p{C20} in
the case $D{=}3$.

The duality for the $\cN{=}8$ vector multiplet can be formulated in
the $v$-analytic superspace.
It is not difficult to define the triplet of $v$-analytic  superfields
which is dual to the superfield $W^{\s(0,1,1)}$
\bea
&&\omega_v=(\omega^{\s(0,1,-1)}_v,\omega^{\s(0,-1,1)}_v,
\omega^{\s(0,0,0)}_v)~,\lb{E22}\\
&&D^{\s(\pm1,1,0)}\omega_v=D^{\s(\pm1,0,1)}\omega_v=D_c^{\s(\pm2,0,0)}
\omega_v=0~.\lb{E23}
\eea
These superfields have an infinite number of auxiliary components.

The interpolating term for the duality relation has the following form:
\bea
&&\int dv dt D^{\s(0,-2,-2)}
[W^{\s(0,1,1)}D_v^{\s(0,1,1)}\omega^{\s(0,0,0)}_v\nn\\
&&+W^{\s(0,1,1)}D_v^{\s(0,2,0)}\omega^{\s(0,-1,1)}_v +
W^{\s(0,1,1)}D_v^{\s(0,0,2)}\omega^{\s(0,1,-1)}_v]~,\lb{E24}
\eea
where $W^{\s(0,1,1)}$ is treated as an unrestricted $v$-analytic
superfield.
\vspace{0.5cm}

{\bf Acknowledgments}
\vspace{0.5cm}

I am grateful to E.A. Ivanov for the stimulating discussions.
This work is partially supported  by  grants  RFBR-96-02-17634,
RFBR-DFG-96-02-00180,  INTAS-93-127-ext and INTAS-96-0308, and
by  grant of Uzbek Foundation of Basic Research N 11/97.
\vspace{0.5cm}

{\bf Note added in the replaced version}
\vspace{0.5cm}

In this new e-print version, I have corrected the formulas (2.26-2.29) in 
connection with the recent results on the component non-Abelian 
superconformal-invariant $5D$ action:\\
T. Kugo and K. Ohashi, Prog. Theor. Phys. 105 (2001) 323; hep-ph/0010288.

\end{document}